\documentclass[]{tPHM2e}

\begin{document}
\doi{10.1080/1478643YYxxxxxxxx}
\issn{1478-6443}
\issnp{1478-6435}
\jvol{00} \jnum{00} \jyear{2010} \jmonth{?? June}

\markboth{Taylor \& Francis and I.T. Consultant}{Philosophical Magazine}


\title{NOVEL PROPERTIES OF FRUSTRATED LOW DIMENSIONAL MAGNETS WITH PENTAGONAL SYMMETRY}

\author{A. Jagannathan$^{\rm a}$$^{\ast}$\thanks{$^\ast$Corresponding author. Email: jagannathan@lps.u-psud.fr
\vspace{6pt}}, Benjamin Motz$^{\rm a}$ and E. Vedmedenko$^{\rm b}$\\\vspace{6pt}  $^{\rm a}${\em{Laboratoire de physique des solides, Universite Paris Sud, 91405 Orsay, France}}; $^{\rm b}${\em{Department of Physics, Universit\"{a}t Hamburg, D-20355 Hamburg,
Germany}}\\\vspace{6pt}\received{v4.4 released November ????} }

\maketitle

\begin{abstract}
  \bigskip 

  \noindent
  In the context of magnetism, frustration arises when a
  group of spins cannot find a configuration that minimizes all of
  their pairwise interactions simultaneously. The simplest example is
  that of a triangle in which the three corners are occupied by Ising
  spins, and where each pair is coupled antiferromagnetically. We
  consider a new type of frustrated spin networks with pentagonal
  loops and long range quasiperiodic structural order. Five-fold loops
  can be expected to occur naturally in quasicrystals, and
  experimental studies of icosahedral alloys show manifestations of
  local five-fold symmetry in a variety of different physical
  contexts. Our model considers classical spins placed on vertices of
  a subtiling of the two dimensional Penrose tiling, and interacting
  with nearest neighbors via antiferromagnetic bonds. The ground state
  of this fractal system has a complex magnetic structure, and is studied
  analytically as well by Monte Carlo
  simulation on finite clusters.  We discuss 
  configuration entropy, and the energy of the ground state of this model.

\bigskip

\end{abstract}

\section{Introduction}

Frustrated magnetic models have been a subject of much interest over
the years, due to the new types of behaviors that can
arise as a result of the frustration. Spin glasses are one well-known example of frustrated
magnets. In these materials, the magnetic moments (spins) interact via
exchange interactions that are random in sign and, in most cases, in
magnitude as well. As a consequence, any given spin is typically
unable to find an orientation that can “satisfy” all of the bonds that
join it to its neighbors, whence the idea of “frustration”. Spins
freeze below a certain transition temperature, with randomly oriented
moments, forming a disordered magnetic “glassy” phase, with complex
dynamical and thermodynamical properties.  Periodic frustrated systems
include the Ising antiferromagnet on a triangular lattice, whose interesting zero temperature properties were considered by
Wannier in 1950.  In the triangular antiferromagnet, the frustration has a geometric
origin, due to the fact that one cannot place three vectors on
the three vertices of a triangle so as to satisfy simultaneously all of the antiferromagnetic
bonds. Similarly, in three dimensions, the presence of triangular loops in the face-centered cubic lattice leads to frustration. In recent years much interest has been shown in the quantum states of such frustrated antiferromagnets, and much work has been done on frustrated
systems such as the Kagome and pyrochlore lattices, spin ice models, etc (for a review see
\cite{review}).

\begin{figure}[t] 
\begin{center}\includegraphics[width=0.5\linewidth]{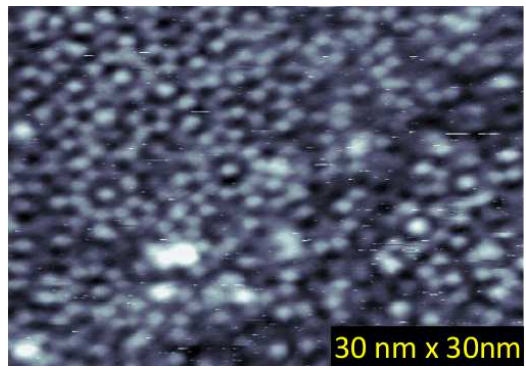}\end{center}
\caption{STM image of Pb atoms deposited on a AgInSb surface showing
the presence of five and ten-fold atomic rings (figure reproduced with
permission, courtesy of the Liverpool and Nancy groups)}
\label{fig:pbonag}
\end{figure}

In this report, we will introduce a new type of frustrated model which
is based on five-fold symmetry, present in many
quasicrystals. Motivation for considering such a quasiperiodic magnet
comes from the experimental progress in recent years in producing
quasiperiodic atomic layers on the surfaces of quasicrystals. Many
different atomic species have been shown to form pseudomorphic films
with a connectivity that reflects the underlying quasiperiodic
organisation as can be seen in Fig. \ref{fig:pbonag} which shows an
STM image of adsorbed Pb atoms on the surface of AlPdMn (for more
details see the review in \cite{sharma}). It is thus likely that
magnetic structures could eventually be obtained as such techniques are refined and
extended. 

As a simple theoreticians model of a frustrated two dimensional quasiperiodic magnetic layer, we
consider antiferromagnetically coupled vector classical spins on
vertices of a subtiling of the Penrose tiling.  We find that the
resulting structure is frustrated on an infinity of length scales,
with, as the starting point the elementary pentagonal five-fold rings.
The frustration leads naturally to a macroscopic entropy of the ground
states of this system, as has been shown for the six-fold symmetric
Kagome lattice antiferromagnet, which is based on the triangular
elementary unit. The quasiperiodic antiferromagnet provides a new
playground for exploration of the physics of frustrated systems.  

We present the model in the following section \ref{sec:model}. The first subsection
outlines some details regarding the geometrical properties of the
model. The following subsection outlines the methods of analysis that
we have used. Section \ref{sec:result} presents results, along with a
discussion and conclusions.

\section{A model of a frustrated quasiperiodic antiferromagnet}
\label{sec:model}

\begin{figure}[t]
\begin{center}\includegraphics[width=0.5\linewidth]{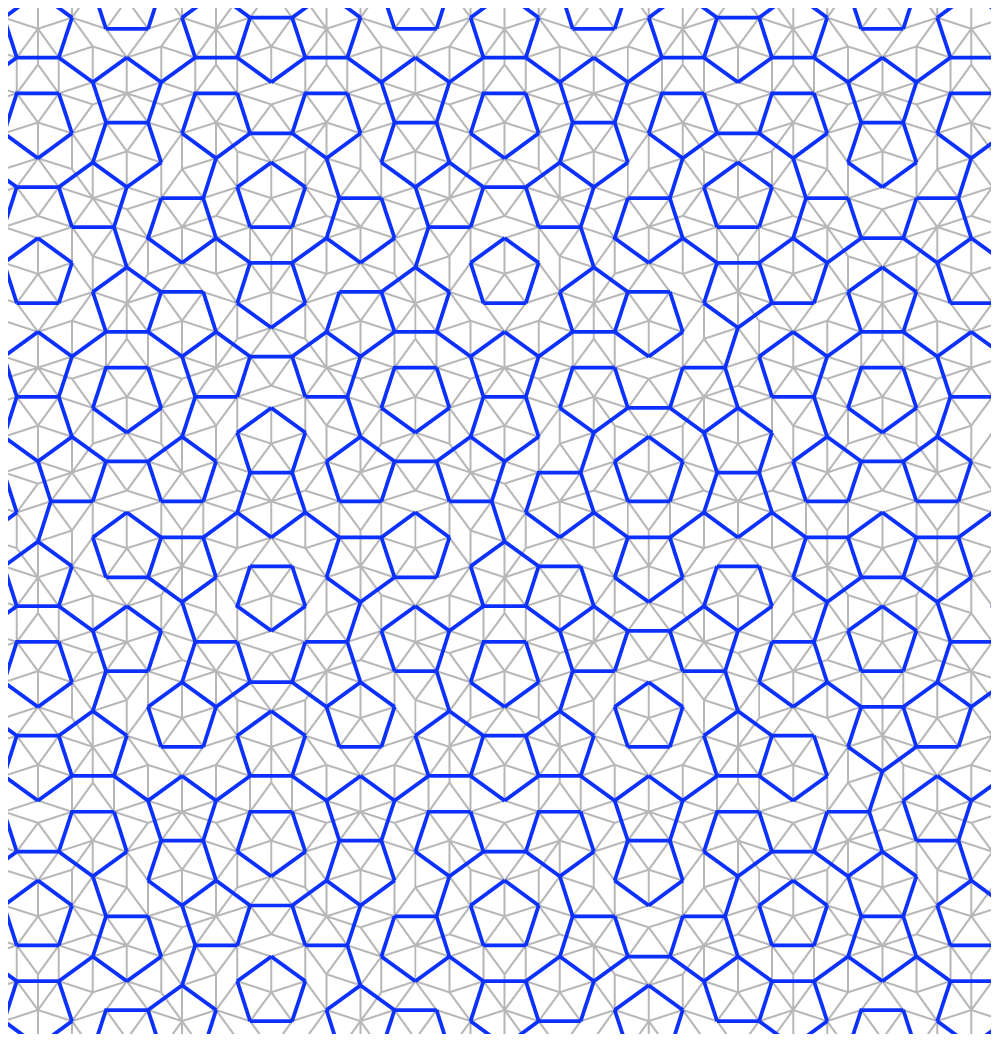}\end{center}
\caption{The vertices of the subtiling, along with the couplings are
shown in blue, with the original Penrose tiling shown in light gray
for comparison.}
\label{fig:tiling}
\end{figure}

We consider two-component vector spins (XY spins) $\vec{S}_i$ placed
on vertices, labeled by the index $i$, of a subtiling of the Penrose
rhombus tiling (PT), as shown in Fig.2. The Hamiltonian is
\begin{eqnarray}
 H = J \sum_{i,j} \vec{S}_i\cdot\vec{S}_j
\end{eqnarray}

where $J$ is the strength of the spin-spin interactions.  All the
couplings are antiferromagnetic ($J>0$), they are of equal strength and
act along the short diagonals of the fat rhombuses, indicated by blue
bonds in Fig. \ref{fig:tiling}.

Spins are clearly frustrated due to the presence of the pentagonal
rings. The ground state degenarcy is easily calculated. In the case of Ising variables, it is five, since there are five ways to have one frustrated bond in a given pentagon. For the vector spin case, the degeneracy is reduced to
just two, as we explain below. However, the frustration is also present at larger length scales, and this is  a consequence of the
geometrical properties of the structure, some of which we outline in the next subsection. This in turn is reflected in the ground state entropy of this frustrated antiferromagnet.

\subsection{Geometrical properties of the pentagonal antiferromagnet
(PAF) }

The system illustrated in Fig. \ref{fig:tiling} is obtained from the Penrose rhombus tiling by occupying a subset of the sites with spins. The figure shows the couplings between pairs of spins which act along the bonds of length $\tau=(\sqrt{5}-1)/2$ (in units in which the edge length of a rhombus is equal to 1). All bonds have the same strength and are antiferromagnetic. 

The PAF is an assembly of interpenetrating clusters of increasing size, which are five-fold symmetric and fractal. They are labeled $C_n$, where $n$ is the ``generation'' of the cluster. The smallest clusters are isolated
pentagons ($C_0$ and $C_1$). Next, there are clusters of five pentagons
arranged around a decagonal ring ($C_2$). The first few members of the
sequence of clusters are shown in
Fig. \ref{fig:smallgroundstates}. Each cluster can be constituted
recursively from three preceding ones. 

\begin{figure}[ht] 
\begin{center}\includegraphics[width=0.7\linewidth]{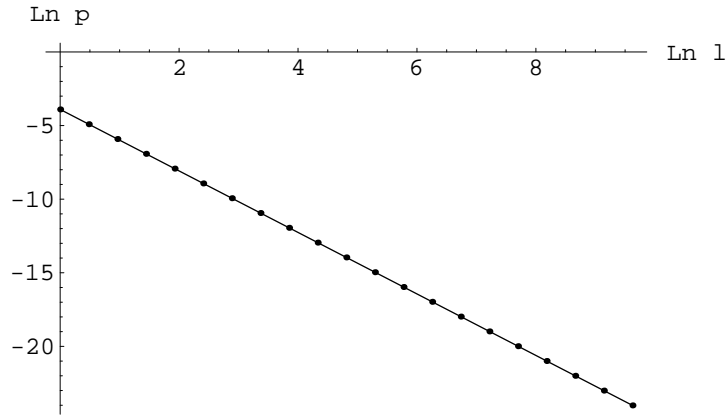}\end{center}
\caption{A log-log plot of the probability $p$ of each cluster plotted against its linear
size $l$, showing the power law dependence. The slope is equal to -2.089. }
\label{fig:powerlaw}
\end{figure}

\begin{figure}[ht] 
\begin{center}\includegraphics[width=0.7\linewidth]{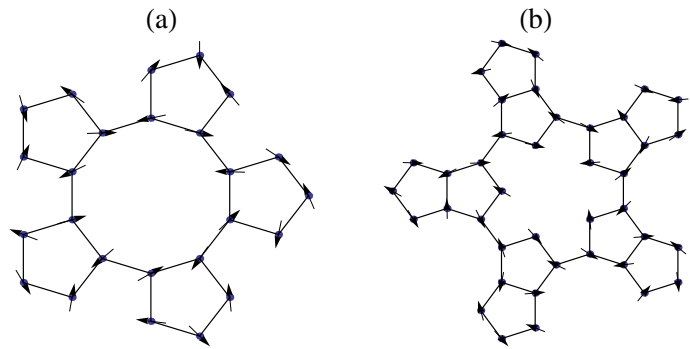}\end{center}
\caption{One of the ground state spin configurations for the
clusters $C_2$ and $C_3$ containing 25 and 40 spins respectively}
\label{fig:smallgroundstates}
\end{figure}

The properties of the infinite
cluster, which percolates through the system can thus be
calculated to great accuracy by a recursive method. We thus find the fractal dimension of the infinite cluster
to be 1.799. The probability of occurrence of each type of cluster, $p_n$,
falls off with the cluster linear size $l_n = \tau^n$ as a power law. Fig. \ref{fig:powerlaw} shows the
distribution obtained numerically for the PAF for large finite samples obtained using the method of
cut-and-projection. The slope gives the exponent of the power law dependence $p \propto l^{-\gamma}$ with $\gamma=2.089$.

\subsection{Magnetic ground states}

For XY spins on the vertices of a pentagon, the minimal energy states
correspond to configurations in which the spins angles are rotated by
$\pm 4\pi/5$ as one passes from one vertex to the next.  The bond energies are all equal, and
have the value $b= J\cos(4 \pi/5)$. For a given pentagon, we can define $\chi = \sum_{bonds} (\phi_i -\phi_j)/4 \pi $ where is
the sum of changes of angle around the five bonds of the pentagon. The overall
change of angle is thus $\pm 4\pi$ and there is a two-fold chiral
degeneracy of the ground state, $\chi =\pm 1$.

\begin{figure}[ht] 
\begin{center}\includegraphics[width=0.7\linewidth]{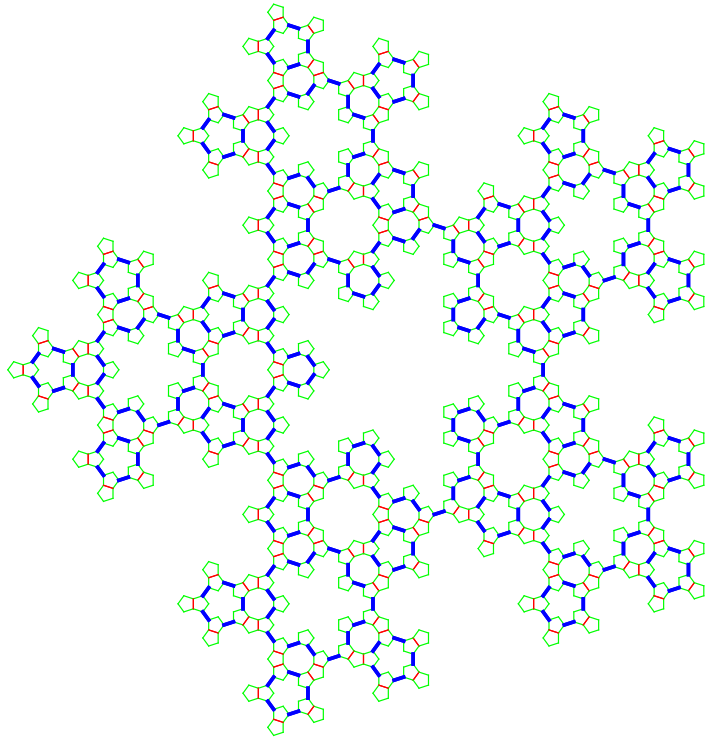}\end{center}
\caption{A cluster of 1515 spins, with the bonds colored according to
the expected corresponding energies in the ground state. Theoretical
predictions show that bonds fall into three categories : the most
frustrated bonds (red), intermediate frustrated bonds (green) and the
least frustrated bonds (blue).}
\label{fig:frustrationclasses}
\end{figure}

The next biggest cluster is composed of 25 spins arranged in five
linked pentagons. The resulting structure is frustrated, both at the
level of the pentagonal rings and, as a result, at the level of the
decagonal loop as well. One can distinguish three types
of bonds: $b_1$, bonds belonging to the pentagons, $b_2$, the bonds that belong both to a pentagon and the decagon and $b_3$, the bonds of the decagon that connect
two different neighboring pentagons. By a direct minimization of the
cluster Hamiltonian, we can easily determine the optimal values of
each of the bond energies. $b_1$ and $b_2$ correspond
to a difference of angles close to $4\pi/5$, while $b_3$ is close
to the optimal value of $-J$, or equivalently to a difference of angles close to $\pi$. These last are the least “frustrated” bonds
in the cluster. Bigger clusters have a greater number of inequivalent bonds, but one finds that the values of bonds belong in one of three main groups. This is shown in fig. \ref{fig:frustrationclasses} which shows a color coded
representation of the main classes of bond energies expected in the
ground state of a large cluster.

Coming now to the question of degeneracy, our analysis shows that the
$C_2$ clusters have many degenerate states. The calculation of the configurational entropy of the 
will be reported elsewhere. We note here simply that this feature is an generalization
of the chiral degeneracy already mentioned for the pentagons. 

\begin{figure}[h!] 
\begin{center}\includegraphics[width=0.5\linewidth]{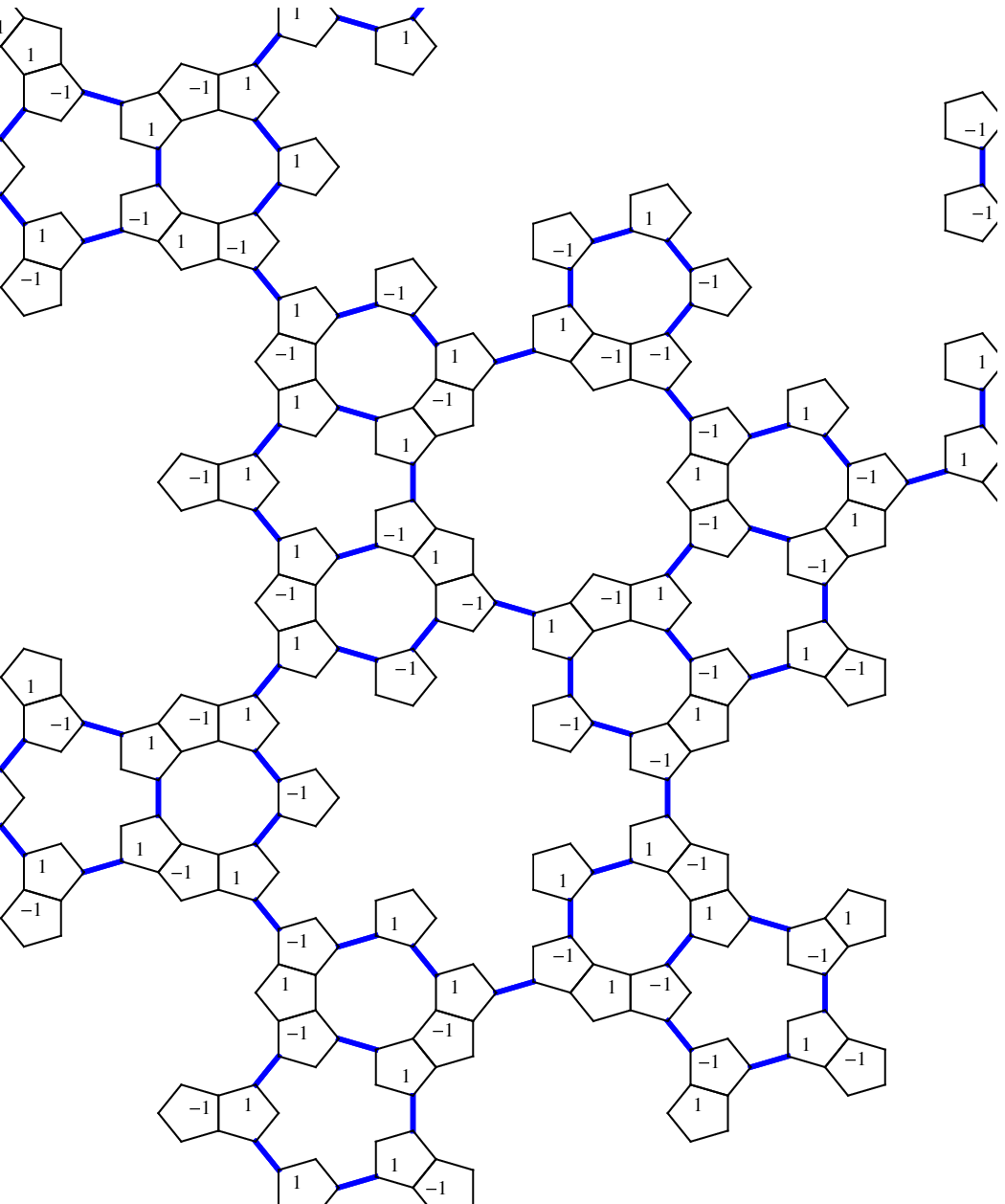}\end{center}
\caption{A small region of a large cluster, representing one of the possible ground state spin
configurations, with the chirality indices shown for each of the
pentagons (Nb. one defect is present in this figure.)}
\label{fig:chiralities}
\end{figure}

\section{Results and Conclusions}
\label{sec:result}
\begin{figure}[ht] 
\begin{center}\includegraphics[width=0.5\linewidth]{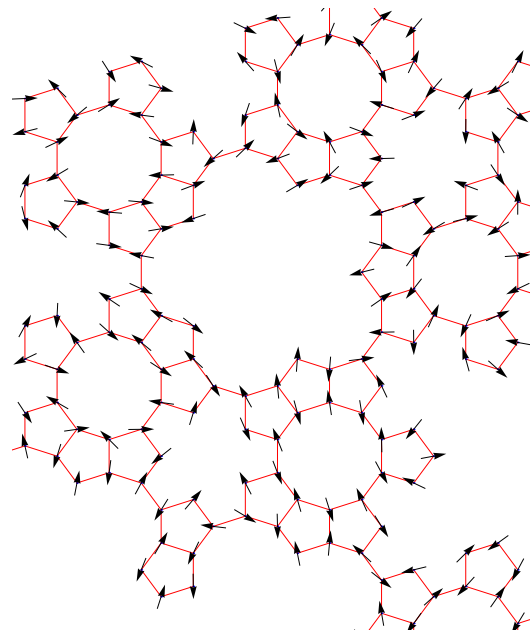}\end{center}
\caption{A portion of a large cluster showing the local spin
orientations as obtained by Monte Carlo simulation (T=0.05J). The agreement
between theory (fig. \ref{fig:frustrationclasses}) and calculations is
found to be good.}
\label{fig:numerics}
\end{figure}

Fig. \ref{fig:numerics} shows the result of a Monte Carlo simulation
of a cluster of 1515 spins. The spin orientations for this sample
follow the theoretically predicted schema. Deviations occur because of the low dimensionality of the
system despite the low temperature (T=0.05 in units of J) used for the simulation. There is a good accord between the analytically and numerically determined states.

\subsection{Conclusions}
We have presented a new class of frustrated antiferromagnet based on
five-fold frustrated units. The structure is fractal, and has characteristic scaling properties in real space as well as perpendicular that follow from the algebraic properties of the underlying Penrose tiling. On the
theoretical side, the subtiling is an interesting geometrical object
in its own right, as it forms a percolating subsystem. It is conceivable that this type of structure could be realized by atomic
deposition on quasiperiodic substrates, in which the adsorbate atoms sit on preferential sites. This would provide an exciting
opportunity for exploring the physics of frustrated magnets.  Problems
we would like to address in the near future concern the spin dynamics and
magnetic response of this type of magnetic material. Studies of the
low energy magnetic states for the long
range dipolar spin-spin interactions considered in \cite{vedmed} are also
in progress, and results will be reported elsewhere. The quantum version of this model, similar to what was discussed for the unfrustrated Penrose tiling case in \cite{attila} will be an interesting theoretical challenge.

\subsection{Acknowledgements}
We would like to thank Pavel Kalugin for useful discussions. Benjamin
Motz was supported by a CNRS contract for the duration of this work.

\end{document}